\documentclass[journal]{IEEEtran}

\usepackage{amsmath}
\usepackage{amssymb}
\usepackage{graphicx}
\usepackage{cite}
\usepackage{url}
\usepackage{comment}
\usepackage{makecell}
\usepackage[safe]{tipa}

\begin{document}

\title{Self-Supervised Source-Filter Model of Speech Production}

\author{ Nani Meka,  Raghavan Ramesh, and Sri Rama Murty Kodukula%
\thanks{This work has been submitted to the IEEE for possible publication. Copyright may be transferred without notice, after which this version may no longer be accessible.}
\thanks{The authors are with the Speech Information Processing Lab, Department of Electrical Engineering, Indian Institute of Technology Hyderabad, India (Corresponding author:  Nani Meka, e-mail: ee23resch01001@iith.ac.in).}}

\maketitle

\begin{abstract}
Deep learning models of speech seldom represent the source-filter process by which speech is produced, relying instead on latent features with no physical correspondence to articulation or phonation. We propose SoFilNet, a self-supervised, differentiable Source--Filter Network that estimates the parameters of speech production directly from the raw waveform: a source, specified by fundamental frequency (F0), voicing and the aperiodicity of five frequency bands, and a filter, specified by ordered formant poles, additional poles that shape the rest of the spectral envelope, and antiresonance zeros. The model is trained only to reconstruct the input speech; no formant or pitch annotation is used. Since per-bin spectral losses favor blurred spectra wherever the model is uncertain, we train F0 and voicing against autocorrelation pseudo-labels refined with the model's own filter, aperiodicity against the periodic fraction of each band, and formant sharpness against spectral contrast. On VTR-FORMANTS, the estimated formants attain the lowest mean error among the reported methods, including supervised trackers, and on the unseen PTDB-TUG corpus the F0 frame error matches the best classical tracker and is lower than that of all learned trackers. The parameters are also consistent with speech production theory: aperiodicity rises with frequency and orders phone classes by the turbulence of their excitation, and the zeros deepen on nasal consonants, the lower one rising with place of articulation.
\end{abstract}


\section{Introduction}
\label{sec:intro}

Speech is produced by a source, the quasi-periodic vibration of the vocal folds or turbulent airflow, shaped by the vocal tract, whose configuration imposes a time-varying set of resonances, the formants, and, when side cavities are coupled in, antiresonances \cite{fant1960acoustic}. In this source-filter view, speech is fully described by two time-varying components. The source is characterized by voicing, F0 and its periodic-to-aperiodic energy ratio across frequency; the filter, by its resonances and antiresonances.

Recovering this description from the acoustic signal is difficult. The source and the filter vary continuously and largely independently: formants move with the articulators, while F0, voicing and voice quality change with prosody, phonation and consonantal context. Both must therefore be recovered jointly from a single signal in which they are entangled. In voiced speech, the envelope of the filter is observed only at the harmonic frequencies of the source, so errors in estimating one propagate into the other. Broad characteristics, such as the overall spectral slope, can be attributed to either. The field has treated the two largely separately, with different methods, assumptions and evaluation conventions.

Formant estimation has proven the harder of the two. Classical tools such as Praat \cite{boersma2002praat} and WaveSurfer \cite{sjolander2000wavesurfer} fit an all-pole linear-prediction model to each frame and pick formants among its poles. They need no training data, but the all-pole assumption cannot represent the antiresonances of nasals and laterals, the envelope is pulled towards individual harmonics when F0 is high, and the assignment of poles to formants relies on heuristics and on speaker-specific settings \cite{schiel2018formant}. Closed-phase analysis \cite{gowda2017qcpfb} reduces the influence of the source, but depends on a well-defined glottal cycle. KARMA \cite{mehta2012karma} tracks formants and antiresonances jointly with a Kalman filter, but relies on a hand-specified state-space model. Supervised neural trackers \cite{dissen2019deepformants} are the most accurate methods available, but need scarce expert-corrected formant tracks and estimate the formants alone. FormantNet \cite{lilley2021formantnet} removes the need for labels by reconstructing a spectral envelope from estimated poles and zeros, but discards the glottal source beforehand by a fixed inverse filtering step, whose errors pass unchecked into the formants.

Pitch estimation is more mature. Classical trackers, from the inverse-filtering SIFT \cite{markel1972sift} to RAPT \cite{talkin1995rapt}, pYIN \cite{mauch2014pyin} and SWIPE \cite{camacho2008swipe}, are fast and reliable on clean speech, but their peak-picking rules make octave errors, their voicing decisions depend on hand-set thresholds, and they suppress the vocal-tract response rather than estimate it. Supervised networks such as CREPE \cite{kim2018crepe} are robust and accurate, but depend on annotated or synthetic training data. Self-supervised trackers avoid annotation altogether. SPICE \cite{gfeller2020spice} and PESTO \cite{riou2025pesto} learn from pitch-shifted copies of the input, whose pitch differs by a known amount; this determines pitch only up to an offset, which must be calibrated separately. Engel et al. \cite{engel2020selfsupervised} instead generate synthetic audio with a differentiable synthesizer, train a network to recover the known synthesis parameters, including pitch, and apply it to real recordings. Its accuracy depends on how well the synthetic signals cover real speech, and the synthesizer describes timbre rather than the vocal tract. Concurrently with this work, Yu and Fazekas \cite{yu2026inverse} track F0 and glottal phase without supervision by inverse voice synthesis; their model includes an all-pole vocal-tract filter, but only the source is evaluated.

The models that do describe both the source and the filter come from speech synthesis and coding. Harmonic-plus-noise models \cite{stylianou2001hnm}, multiband excitation coders \cite{griffin1988mbe}, and STRAIGHT and WORLD \cite{kawahara2001aperiodicity,morise2016world} represent a frame by F0, voicing, a spectral envelope and an aperiodicity measure per frequency band. Their parameters are obtained by separate signal-processing analyses, and the spectral envelope is a smooth curve rather than a set of identified formants. Differentiable digital signal processing (DDSP) \cite{engel2020ddsp} made harmonic-plus-noise synthesis trainable end to end, but typically takes F0 from an external tracker and represents the filter as a free-form envelope.

In this paper we estimate a source-filter description of speech jointly from the raw waveform, with a single network and no annotation. We call the model SoFilNet, for \textbf{So}urce--\textbf{Fil}ter \textbf{Net}work. At every 10\,ms frame, the network predicts a source, specified by F0, voicing and the aperiodicity of five frequency bands, and a filter, specified by ordered formant poles, additional poles that shape the rest of the spectral envelope, and antiresonance zeros. A differentiable synthesizer turns these parameters into the spectrum of the frame, and the model is trained to reconstruct the input speech. Formant identity is guaranteed by construction, so no tracking or labeling step is needed.

Without labels, the model can only be trained by comparing its predicted spectrum with the observed one, typically frequency bin by frequency bin. Such per-bin comparisons have a systematic blind spot. They penalize a sharp spectral peak placed slightly away from its true frequency more heavily than a broad, flat peak covering the same region, and they penalize a noise-like spectrum, whose random fine structure can never be matched bin by bin, more heavily than a smoother, partly harmonic one. Wherever the model is uncertain, the loss favors a blurred spectrum over a sharp one. This single property accounts for three problems we encountered. While F0 is wrong, the loss prefers a smooth source to a misplaced harmonic comb, so F0 receives little useful gradient and is learned poorly, a known weakness of spectral distances for pitch \cite{turian2020sorryforyourloss}. Aperiodicity is underestimated, because a partly harmonic source fits better bin by bin than genuine noise. Formants come out too broad, with the valleys between them too shallow.

We therefore train each of these quantities against a target that does not depend on the exact position of individual spectral peaks. F0 and voicing are supervised by pseudo-labels derived from the autocorrelation of the input. These are obtained first from the speech itself, and later from its residual after inverse filtering with the model's own vocal-tract estimate. The targets thus improve as the filter improves, much as the pseudo-labels of HuBERT are refined with the model's own representations \cite{hsu2021hubert}. Aperiodicity is trained through a proxy: the fraction of periodic energy in each frequency band of the predicted spectrum is matched to that of the speech, and this fraction falls steadily as aperiodicity grows. Formant sharpness is likewise trained through the spectral contrast of each band \cite{jiang2002contrast}, the level difference between its strongest and weakest components. Our contributions include:
\begin{itemize}
\item A differentiable source-filter model, estimated from the raw waveform, with band aperiodicity in the source and formants ordered by construction in the filter.
\item Three training targets that remedy the blind spot of per-bin losses without annotation: refined autocorrelation pseudo-labels for F0 and voicing, a band periodic-fraction loss for aperiodicity and a spectral-contrast loss for formant sharpness.
\item Formant tracking on VTR-FORMANTS with the lowest mean error among the reported methods, including supervised trackers, and F0 tracking on an unseen corpus, PTDB-TUG, with an F0 frame error that matches the best classical tracker and is lower than that of all learned trackers.
\item Evidence that the individual parameters behave as speech production theory predicts: band aperiodicity rises with frequency and orders phone classes by the degree of turbulent excitation, and the zeros deepen on nasal consonants, the lower one rising with place of articulation.
\end{itemize}

\section{Differentiable Source-Filter Model}
\label{sec:source-filter}

Speech is analyzed in short frames, over which the glottal source and the vocal tract can be treated as stationary. Within a frame, the speech spectrum is the product of the source spectrum and the frequency response of the vocal tract, $S(j\omega) = G(j\omega)\,V(j\omega)$ \cite{fant1960acoustic}. We parameterize the source as a harmonic-plus-noise mixture \cite{stylianou2001hnm}, described by $\lambda_g$: voicing, F0 and an aperiodicity in each of five frequency bands \cite{kawahara2001aperiodicity,zen2007nitech}. We parameterize the vocal tract as a pole-zero filter, described by $\lambda_v$: the frequencies and bandwidths of its resonances and antiresonances. A per-frame gain $\gamma$ sets the overall level. The predicted spectrum is a closed-form, differentiable function of $\lambda = \{\lambda_g, \lambda_v, \gamma\}$, so it can be compared directly with the observed spectrum and the error propagated back to the network that estimates $\lambda$.

\subsection{Harmonic-plus-noise model of the glottal source}
\label{sec:harmonic-noise}

When the vocal folds vibrate, the airflow is chopped into quasi-periodic pulses, giving a harmonic excitation with energy at multiples of F0. When they are apart, or vibrate without full closure as in breathy phonation, the airflow becomes turbulent, giving a broadband aperiodic excitation. Voiced speech generally contains both, in proportions that vary with frequency and with phonation. We therefore model the source as a mixture of a harmonic and an aperiodic component.

\subsubsection*{Harmonic component}
With F0 normalized as $\theta_0 = 2\pi F_0/f_s$, $f_s$ the sampling frequency, a comb of $K$ equal-amplitude harmonics below the Nyquist frequency has the closed form
\begin{equation}
g_h[n] = \sum_{k=1}^{K}\cos(k\theta_0 n) = \frac{\sin(K\theta_0 n/2)}{\sin(\theta_0 n/2)}\cos\!\left(\frac{(K+1)\,\theta_0 n}{2}\right),
\label{eq:harmonic-closed}
\end{equation}
whose evaluation cost does not depend on $K$, so frames with different F0 are synthesized in parallel. Its spectrum is
\begin{equation}
G_h(j\omega) = \sum_{n=0}^{L-1} w[n]\,g_h[n]\,e^{-j\omega n},
\label{eq:harmonic-fd}
\end{equation}
where $w[n]$ is the Hann window of length $L$ used to analyze the speech itself, so that a correctly estimated comb has exactly the peak shape of the real harmonics.
\subsubsection*{Aperiodic component}
Turbulent airflow is modeled as stationary white Gaussian noise $g_a[n]$, whose power in each frequency bin is exponentially distributed. We therefore draw $|G_a(j\omega)|^2$ independently in each bin from a unit-mean exponential distribution, so that the noise contributes the irregular fine structure of a real noise spectrum. The share of the noise power varies with frequency, depending on the sound and the phonation. To capture this, an aperiodicity $a_b \in [0,1]$ is assigned to each of five bands $b$: 0--1, 1--2, 2--4, 4--6 and 6--8\,kHz, the bands used in STRAIGHT-based synthesis \cite{zen2007nitech}, and the values are interpolated linearly between the band centers to obtain a smooth curve $a(\omega)$. The aperiodic part of a voiced frame is the noise spectrum scaled by this curve, $a(\omega)\,|\hat{G}_a(j\omega)|^2$, where $\hat{G}_a$ is $G_a$ normalized to unit energy.
\subsubsection*{Mixture}
The harmonic part of a voiced frame receives the complementary weight, $\big(1 - a(\omega)\big)\,|\hat{G}_h(j\omega)|^2$, with $\hat{G}_h$ the comb normalized to unit energy, so that at every frequency $a(\omega)$ is the share of the noise in the mixing weights. A voiceless frame is excited by noise alone. With the voicing decision $u \in \{0,1\}$, the source spectrum is modeled as
\begin{equation}
\begin{split}
|G(j\omega)|^2 = {} & u\Big[\big(1 - a(\omega)\big)|\hat{G}_h(j\omega)|^2 + a(\omega)\,|\hat{G}_a(j\omega)|^2\Big]\\
& + (1-u)\,|\hat{G}_a(j\omega)|^2.
\end{split}
\label{eq:source-mix}
\end{equation}
The two components are added in power because they arise from independent processes.
Both components have a flat expected spectrum, so the source carries no spectral slope of its own; the slope of the glottal pulse is left to the filter, as in linear prediction \cite{atal1971speech}. The source parameters are $\lambda_g = \{F_0, u, a_1, \dots, a_5\}$.

\subsection{The filter: a pole-zero cascade}
\label{sec:filter}

The vocal tract is modeled by a rational transfer function, the classical acoustic-tube approximation \cite{fant1960acoustic,atal1971speech}. Its poles produce the resonances of the tract, the formants, and its zeros produce antiresonances, which arise when the vocal tract branches, as it does in nasals and laterals. Since the filter has real coefficients, its complex poles and zeros occur in conjugate pairs. With $M$ pole pairs $p_k = r_k e^{j\theta_k}$ and $N$ zero pairs $q_k = \rho_k e^{j\phi_k}$, the transfer function is
\begin{equation}
V(z) = \frac{\prod_{k=1}^{N} (1 - q_k z^{-1})(1 - q_k^* z^{-1})}{\prod_{k=1}^{M} (1 - p_k z^{-1})(1 - p_k^* z^{-1})}.
\label{eq:vz}
\end{equation}
Its squared magnitude response on the unit circle follows in closed form,
\begin{equation}
|V(j\omega)|^2 = \frac{\prod_{k=1}^{N} D(\omega;\rho_k,\phi_k)}{\prod_{k=1}^{M} D(\omega;r_k,\theta_k)},
\label{eq:filter}
\end{equation}
where $D(\omega; r, \theta) = \big(1 + r^2 - 2r\cos(\omega - \theta)\big)\big(1 + r^2 - 2r\cos(\omega + \theta)\big)$ is the squared magnitude of one conjugate pair. Each pole pair produces a peak at $\omega = \theta_k$ whose bandwidth is approximately $B_k = -(f_s/\pi)\ln r_k$ in Hz, so that a radius close to one gives a sharp resonance. Each zero pair produces a notch at $\omega = \phi_k$ that deepens as $\rho_k$ approaches one and vanishes as $\rho_k$ approaches zero. Only the shape of $V$ matters, since it is normalized before being combined with the source, so no filter gain is parameterized. The filter parameters are $\lambda_v = \{\{r_k, \theta_k\}_{k=1}^{M}, \{\rho_k, \phi_k\}_{k=1}^{N}\}$.

\subsection{Combining source and filter}
\label{sec:combining}

\begin{figure}[t]
\centering
\includegraphics[width=\columnwidth]{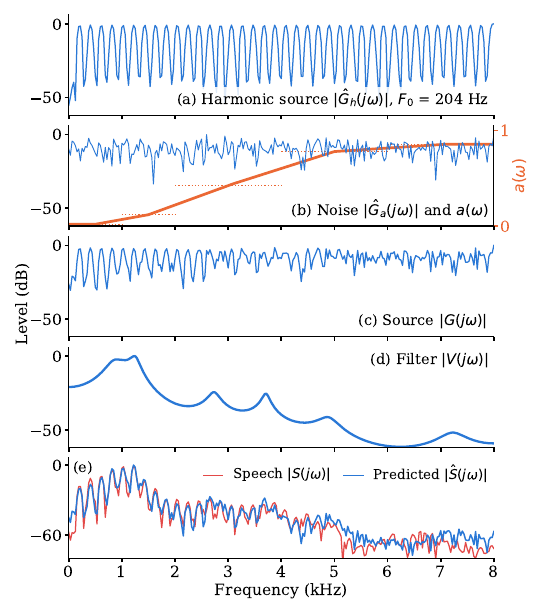}
\caption{Construction of the predicted spectrum for one frame (TIMIT test/dr1/faks0/sx223, /\textipa{A}/ at 2.63\,s), from the parameters estimated by the trained model. (a) Harmonic comb at the estimated F0. (b) Noise spectrum and the aperiodicity curve $a(\omega)$ interpolated from the five band values (dotted). (c) Source spectrum, Equation~\eqref{eq:source-mix}. (d) Filter response, Equation~\eqref{eq:filter}. (e) Predicted spectrum, Equation~\eqref{eq:combine}, and the spectrum of the speech. Panels (a)--(d) are normalized to a 0\,dB maximum.}
\label{fig:sf-illustration}
\end{figure}

The source and filter spectra of Equations~\eqref{eq:source-mix} and~\eqref{eq:filter} are normalized to unit energy and then multiplied. The product is renormalized before a per-frame gain $\gamma$ is applied to set the level of the synthesized speech spectrum:
\begin{equation}
|\hat{S}(j\omega)| = \gamma\,\frac{\tilde{V}(j\omega)\,\tilde{G}(j\omega)}{\|\tilde{V}\tilde{G}\|_2},\quad
\tilde{V} = \frac{|V|}{\|V\|_2},\quad \tilde{G} = \frac{|G|}{\|G\|_2},
\label{eq:combine}
\end{equation}
where $\|\cdot\|_2$ is the Euclidean norm over frequency. Hence the level of the estimated spectrum is determined by $\gamma$ alone, while $\lambda_g$ and $\lambda_v$ determine the shape of the spectrum. For every frame, the network predicts $\lambda = \{\lambda_g, \lambda_v, \gamma\}$, and the predicted spectrum $\hat{S}$ is compared with the ground truth spectrum $S$ to update the network parameters. 
Figure~\ref{fig:sf-illustration} illustrates the steps involved in creating the predicted spectrum from the estimated parameters for a voiced frame.
For this frame, the estimated source is periodic at low frequencies and increasingly noisy above 2\,kHz. When combined with the estimated filter, the resulting spectrum closely matches that of the ground truth speech frame.

\section{Model Architecture}
\label{sec:architecture}

\begin{figure*}[!t]
\centering
\includegraphics[width=\textwidth]{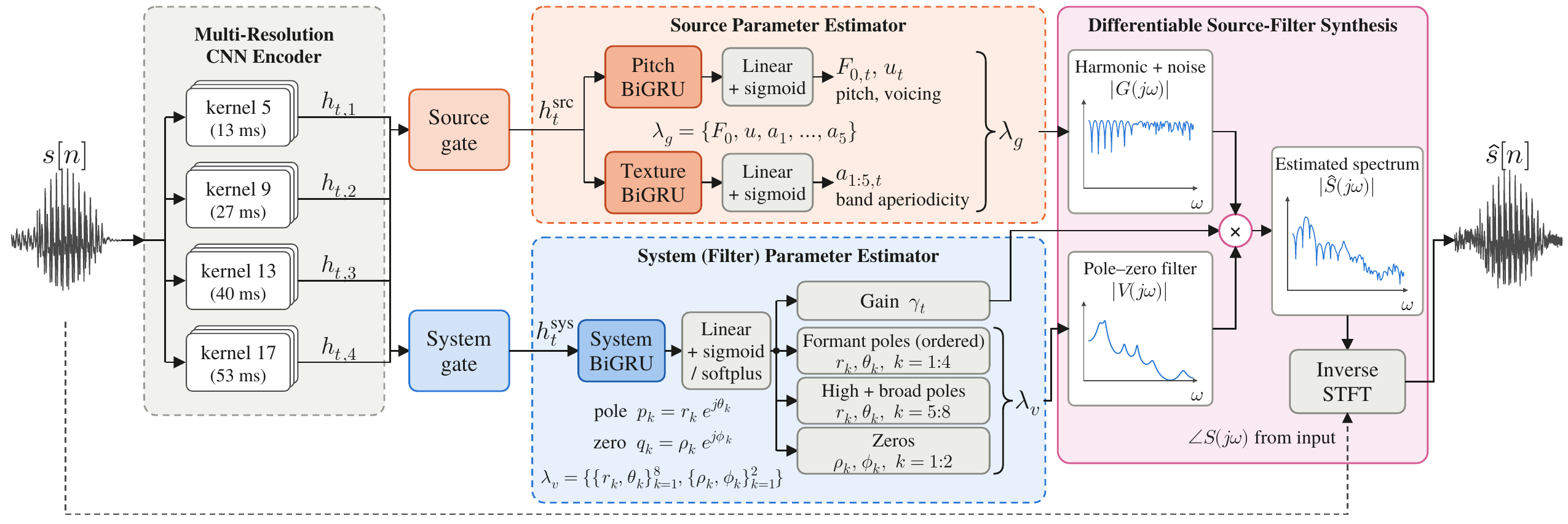}
\caption{SoFilNet architecture. A multi-resolution convolutional encoder extracts features at four receptive fields, which a source gate and a system gate combine per frame. The source parameter estimator (pitch and texture BiGRUs) predicts F0 $F_{0,t}$, voicing $u_t$ and the aperiodicity of five bands, forming $\lambda_g$; the system parameter estimator predicts the gain $\gamma_t$, four ordered formant poles, two poles above the formants, two broad extra poles and two zeros, forming $\lambda_v$. The differentiable synthesizer combines the source spectrum $|G(j\omega)|$ and the filter response $|V(j\omega)|$ into the estimated spectrum $|\hat{S}(j\omega)|$ (Equation~\eqref{eq:combine}); the waveform $\hat{s}[n]$ is reconstructed with the phase of the input for the time-domain losses.}
\label{fig:architecture}
\end{figure*}

The parameters $\lambda = \{\lambda_g, \lambda_v, \gamma\}$ are estimated directly from the raw waveform by a single convolutional-recurrent network, trained end-to-end. Convolutional layers extract local, short-time acoustic features, the learned counterpart of the fixed analysis window of Section~\ref{sec:source-filter}.
Recurrent layers then integrate these features across frames, so the resulting parameter tracks vary smoothly over time, consistent with the continuous articulatory and laryngeal movements that produce them. Figure~\ref{fig:architecture} shows the SoFilNet architecture together with the differentiable source-filter model that follows it.

\subsection{Multi-resolution encoder}
\label{sec:encoder}

When a speech analyst inspects a spectrogram by eye, a single window length is rarely enough. A short analysis window resolves the formant transitions driven by vocal-tract dynamics, but not the harmonic structure of the glottal source, which requires a window spanning several pitch periods. A long window reveals the harmonics, at the cost of smearing the same formant transitions. An analyst switches between the two depending on what they need to see, since no single window serves both purposes. This observation motivates a multi-resolution encoder, which operates at several window lengths in parallel and adaptively selects a combination of them, per frame and per parameter group, for the task at hand.

The multi-resolution encoder consists of four parallel convolutional feature extractors, each with a different receptive field but the same frame rate, so their outputs are temporally aligned. Each is a 4-layer strided convolutional stack with 32, 64, 128, and 256 channels per layer. The four layers use strides of 4, 4, 2, and 5, whose product fixes the output frame rate at 10\,ms regardless of kernel size. In this work we use kernel sizes of 5, 9, 13, and 17 for the four stacks, covering receptive fields from 13 to 53\,ms (Figure~\ref{fig:architecture}). This range is short enough at one end to resolve rapid formant transitions, and long enough at the other to span multiple pitch periods even for the lowest F0 speech encountered. At each frame, two independent softmax gates adaptively combine features from the four resolutions, one feeding the source decoder and the other the system decoder. The gates are kept separate because the two parameter groups favor different resolutions; a single shared gate would impose one compromise window on both, the very trade-off the encoder is designed to avoid.

\subsection{Source decoder}
\label{sec:source-decoder}

The source decoder consists of two parallel bidirectional gated recurrent units (GRUs), a pitch GRU and a texture GRU, each feeding its own linear head mapped to a bounded, physically meaningful range (Figure~\ref{fig:architecture}). Separate recurrent stacks keep the F0 and voicing objectives from competing with the texture parameters for the same recurrent state. The pitch GRU produces voicing and F0. Voicing is a binary classification problem. The head estimates the posterior probability that a frame is voiced, hard-thresholded to the flag $u$ that switches the harmonic component in the source mixture of Equation~\eqref{eq:source-mix}. Since the threshold passes no gradient, the voicing head is trained by its own loss (Section~\ref{sec:losses}). F0, the fundamental frequency of the harmonic component, is regressed using a sigmoid interpolation scaled to the adult pitch range of 60--400\,Hz.

The texture GRU captures the finer characteristics of the voiced regions: the five band aperiodicities $a_1,\dots,a_5$ of Section~\ref{sec:harmonic-noise}, each regressed using a plain sigmoid in the range $[0,1]$, which set the share of periodic and aperiodic energy in each band. Together, F0, voicing and the band aperiodicities form the source parameters $\lambda_g$ used in rendering the source spectrum $|G(j\omega)|$ of Equation~\eqref{eq:source-mix}.

\subsection{System decoder}
\label{sec:system-decoder}

The system decoder estimates the parameters of the pole-zero filter $\lambda_v$ and gain $\gamma$, together approximating the spectral envelope of the speech signal. Multi-resolution features combined by the system gate feed a single bidirectional GRU, whose output drives the pole-zero parameter estimator, as shown in Figure~\ref{fig:architecture}. The total number of poles $M$ of Equation~\eqref{eq:vz} is divided into three groups. $M_F$ formants are constrained to have sharp bandwidths, $M_H$ high poles lie above the highest formant, and the remaining $M_E = M - M_F - M_H$ are extra poles with wide bandwidths, aiding in approximating the overall spectral envelope. The decoder also produces the $N$ zeros of Equation~\eqref{eq:vz} and the gain, whose ranges are described next.

Formants are conventionally identified by their order, $F_1 < F_2 < \dots$, and this ordering is what gives each pole a consistent physical identity across frames. Without an explicit ordering constraint, assigning a consistent identity to each estimated pole would require a fragile, post-hoc tracking step, one that can fail whenever two poles cross or come close in frequency. We circumvent this issue by ensuring consistent ordering of the estimated formants using a stick-breaking recursion. In this approach each formant is estimated using a sigmoid unit, whose range is bounded between the previous formant estimate and a per-formant ceiling of 1.5, 3, 4 and 5\,kHz for F1 to F4. This strategy ensures that the formant ordering, $F_1 < F_2 < \dots < F_{M_F}$, holds at every frame. The pre-sigmoid logit is softly bounded via $6\tanh(x/6)$ rather than hard-clamped. A hard clamp has exactly zero gradient outside its range and can trap a formant that overshoots early in training with no way back, whereas the smooth bound keeps a nonzero gradient that aids recovery. Formant radius is a sigmoid interpolation between a per-formant floor, corresponding to maximum bandwidths of 600, 600, 750 and 750\,Hz for F1 to F4, and the shared ceiling $r_{\max}=0.9999$, keeping each formant a resonance of bounded bandwidth while ensuring filter stability.

High poles model the upper spectrum, where a fifth formant may or may not be present. Each is a sigmoid interpolation between the highest formant and 7.8\,kHz, independently of the other, and its radius is unconstrained. A high pole can thus become a sharp resonance where the spectrum has one and broaden into a gentle envelope shape where it does not, without forcing a formant where none exists. High poles are not ordered among themselves and are not evaluated as formants.

Extra poles (the remaining $M_E$ poles) use plain sigmoid interpolation over their frequency and radius ranges. They are unordered and range freely over 50 to 7800\,Hz, but their radius is upper bounded rather than floored, keeping their bandwidths above 1\,kHz so they shape the residual envelope without being sharp enough to impersonate a formant. Extra poles are thus separated from the formants by bandwidth, not frequency range; an extra pole may sit anywhere, including inside the formant range.

Zeros ($N$ of them) likewise use plain sigmoid interpolation. Their frequency is confined to 250--4000\,Hz, the range of the antiresonances of nasal and lateral consonants \cite{fujimura1962nasal, stevens1998acoustic}; without this restriction, the zeros were used to shape the envelope near 5\,kHz together with the high poles rather than to represent antiresonances. Their radius is unconstrained, with no floor. As the radius of a zero $\rho_k \to 0$, its contribution to the system response approaches unity. So, the model can weaken or switch off a zero in frames that do not require an antiresonance, such as many vowel frames. This is consistent with acoustic theory, in which antiresonances arise from specific articulatory configurations, mainly involving nasals and laterals \cite{fant1960acoustic}.

Finally, gain is regressed using a softplus activation function, giving a positive, unbounded value with a smooth gradient near zero.

Together, the source and system decoders produce the full parameter set $\lambda = \{\lambda_g, \lambda_v, \gamma\}$ at every frame, from which the synthesizer of Section~\ref{sec:source-filter} computes the spectrum $|\hat{S}(j\omega)|$. The network is trained end-to-end by comparing this spectrum, and the waveform reconstructed from it, with those of the input speech, using the loss functions described in the next section.

\section{Loss Functions for Model Training}
\label{sec:losses}

The proposed model is an autoencoder with an interpretable bottleneck. In place of a learned latent space, the network encodes speech into the parameter set $\lambda$: voicing, F0, band aperiodicity, formants and gain, each with a defined role in the acoustic theory of speech production~\cite{fant1960acoustic,stevens1998acoustic}. The differentiable source-filter model then decodes these parameters back into speech. Since the output of the full network is the reconstructed speech rather than the parameters themselves, it can be trained from raw waveforms alone, without ground-truth parameter values. The model is therefore trained in a self-supervised manner, and the loss functions are the only mechanism that ties each output to its intended physical meaning, rather than to some other decomposition that reconstructs the signal equally well. Accordingly, we use a combination of loss functions that together encourage a faithful reconstruction of the input and keep each estimated parameter tied to its physical meaning. They fall into three groups. Frequency-domain losses compare the estimated and true spectra, bin by bin or band by band. Time-domain losses compare quantities tied to the waveform, including those that train F0 and the voicing decision. Regularization losses constrain the network's internal behavior rather than comparing against the input.

For a signal $x[n]$, we write $x_t[n]$ for its $t$-th frame and $X_t[k]$ for the magnitude of its short-time Fourier transform at frequency bin $k$, the sampled counterpart of $|X(j\omega)|$ in Section~\ref{sec:source-filter}. Thus $s_t[n]$ and $S_t[k]$ refer to the input speech and $\hat{s}_t[n]$ and $\hat{S}_t[k]$ to their estimates, while $S$ and $\hat{S}$ without indices denote the full magnitude spectrograms. Each loss is written for a single frame, and band where applicable, and averaged over frames, bands and the utterances in a batch.

\subsection{Frequency-domain losses}
\label{sec:freq-losses}

\subsubsection{Spectral convergence loss}
Spectral convergence loss measures the relative distance between the estimated and target magnitude spectra, normalized by the norm of the target spectrum. It is scale-invariant and primarily focuses on matching high-energy regions like spectral peaks, which typically correspond to the resonances of the vocal-tract filter and the harmonic peaks of the glottal source.
It uses the Frobenius norm $\|\cdot\|_F$ over the spectrograms of the whole batch rather than frame by frame, so that low-energy bins and frames do not dominate. The spectral convergence loss between the estimated and target spectrograms $\hat{S}$ and $S$ is defined as:
\begin{equation}
\mathcal{L}_{\text{sc}} = \frac{\|\hat{S} - S\|_F}{\|S\|_F}.
\label{eq:l-conv}
\end{equation}

\subsubsection{Spectral log-magnitude loss}
The spectral log-magnitude loss measures the elementwise $L_1$ distance between the logarithms of the magnitude spectra. By compressing large values, it amplifies numerical errors in low-energy regions, forcing the model to fix details in the spectral valleys and silence regions. The log-magnitude loss between the estimated and target spectra of frame $t$, $\hat{S}_t[k]$ and $S_t[k]$, is defined as:
\begin{equation}
\mathcal{L}_{\text{spec}} = \mathbb{E}_{k}\Big[\,\big|\log\hat{S}_t[k] - \log S_t[k]\big|\,\Big]
\label{eq:l-spec}
\end{equation}

\subsubsection{Subband log energy loss}
Comparing individual spectral bins has a specific disadvantage in noise-dominated regions. In such bins, both the estimate and target are different realizations of a random process, so the per-bin comparison is not meaningful. Instead, we compare statistics of the local spectrum over 1\,kHz bands with 50\% overlap. For a band $b$ spanning a set of bins $k \in b$, we define the log energy of a frame restricted to that band as:
\begin{equation}
E_b(X_t) = \log\!\Big(\frac{1}{|b|}\sum_{k \in b}X_t^2[k]\Big)
\label{eq:band-e-def}
\end{equation}
where $X_t$ stands for either $S_t$ or $\hat{S}_t$. The subband log energy loss between the estimated and target spectra in band $b$ is defined as:
\begin{equation}
\mathcal{L}_{\text{en}} = \Big|E_b(\hat{S}_t) - E_b(S_t)\Big|
\label{eq:l-energy}
\end{equation}

\subsubsection{Band periodic-fraction loss}
The band energy fixes the level of each band but not how that energy is divided between harmonics and noise, which carry the same energy at any aperiodicity. Although the per-bin losses do depend on this division, they favor a smoother spectrum in noise-dominated bins, which drives the band aperiodicities $a_b$ below their true values. We therefore compare the fraction of each band's energy that is periodic, following the autocorrelation measure of~\cite{boersma1993hnr}. For a frame with power spectrum $X_t^2[k]$, the autocorrelation of its component in band $b$ at lag $\ell$ is
\begin{equation}
R_b(\ell) = \Big|\sum_{k \in b} X_t^2[k]\, e^{\,j 2\pi k \ell / K}\Big|,
\label{eq:band-acf}
\end{equation}
where $K$ is the DFT length, and its periodic fraction is
\begin{equation}
\rho_b(X_t) = \frac{R_b(\ell^\ast)}{R_b(0)\, w(\ell^\ast)},
\label{eq:periodic-fraction}
\end{equation}
where $w$ is the normalized autocorrelation of the analysis window, which corrects for its taper~\cite{boersma1993hnr}. The lag $\ell^\ast$ is the period of the frame itself: the strongest peak, in the 60--400\,Hz range, of the autocorrelation of its 0--1.5\,kHz component. It is found separately for the input and for the estimate, so the target does not depend on the estimated F0. The fraction is 1 for a perfectly periodic band and falls monotonically towards 0 as the share of aperiodic energy grows. The loss is
\begin{equation}
\mathcal{L}_{\text{pf}} = \Big|\rho_b(\hat{S}_t) - \rho_b(S_t)\Big|
\label{eq:l-pf}
\end{equation}
over the five aperiodicity bands of Section~\ref{sec:harmonic-noise}. It is computed with a 50\,ms window, which holds several pitch periods even at low F0. At that resolution the estimate is rebuilt with every parameter except aperiodicity held fixed, so that $\mathcal{L}_{\text{pf}}$ trains aperiodicity alone; aperiodicity in turn receives no gradient from the other spectral losses.

\subsubsection{Spectral contrast loss}
Per-bin losses favor broad peaks and shallow valleys. When the exact position of a spectral peak is uncertain, a smeared peak lowers the expected per-bin error more than a sharp peak at a slightly wrong frequency. Spectral contrast~\cite{jiang2002contrast} measures the dynamic range of a band, independently of where its peaks lie. Within band $b$, let $P_b(X_t)$ and $Q_b(X_t)$ be the means of the largest and smallest 20\% of the log-magnitudes $\log X_t[k]$, $k \in b$. The contrast of the band and the corresponding loss are
\begin{equation}
C_b(X_t) = P_b(X_t) - Q_b(X_t), \quad \mathcal{L}_{\text{ct}} = \Big|C_b(\hat{S}_t) - C_b(S_t)\Big|,
\label{eq:l-contrast}
\end{equation}
computed over the same bands as $\mathcal{L}_{\text{en}}$. Since it depends only on the dynamic range of each band, this loss rewards sharper formant peaks, that is, narrower formant bandwidths, and deeper valleys between formants and between harmonics. A companion envelope-extrema loss matches the local maxima $U_t[k]$ and minima $D_t[k]$ of $\log X_t$ over five bins centered on $k$ (about $\pm 60$\,Hz),
\begin{equation}
\mathcal{L}_{\text{ex}} = \tfrac{1}{2}\,\mathbb{E}_{k}\big[|\hat{U}_t[k] - U_t[k]| + |\hat{D}_t[k] - D_t[k]|\big],
\label{eq:l-extrema}
\end{equation}
so that a sharp peak slightly misplaced is not penalized for being sharp.

Together, these terms define the frequency-domain loss
\begin{equation}
\mathcal{L}_{\text{freq}} = \mathcal{L}_{\text{spec}} + 0.5\,\mathcal{L}_{\text{sc}} + 0.1\,\mathcal{L}_{\text{en}} + \mathcal{L}_{\text{pf}} + \mathcal{L}_{\text{ct}} + 0.5\,\mathcal{L}_{\text{ex}}
\label{eq:l-freq}
\end{equation}
Except for the per-bin losses, these losses, like the waveform losses below, use only the frames voiced according to the speech pseudo-label.

\subsection{Time-domain losses}
\label{sec:time-losses}

The frequency-domain losses above cannot reliably train voicing and F0. Voicing is a discrete decision gating the source mixture, not a continuous shape these losses can fit, so it needs its own classification signal. F0 is harder still: a per-bin spectral distance responds to a change in F0 only while the predicted harmonics lie close to the true ones, so once they drift apart its gradient no longer indicates which way to move F0 \cite{turian2020sorryforyourloss}. While F0 is wrong, the per-bin losses prefer a smooth source to a misplaced harmonic comb, rather than guiding F0 towards the true value. We give both quantities a direct target derived from the autocorrelation of the signal, and add waveform losses that compare the reconstruction with the input sample by sample.

\subsubsection{Pitch and voicing pseudo-label losses}
The autocorrelation function of a periodic signal peaks at the lag of its period, and that of a noise-like signal does not. Each frame is analyzed with a 50\,ms Hann window, long enough to hold several periods even at low F0, and its autocorrelation is normalized and corrected for the window taper as in Equation~\eqref{eq:periodic-fraction}. The candidate periods are the local maxima of this normalized autocorrelation $r_t[\ell]$ at lags $\ell \in [f_s/F_{0,\max},\, f_s/F_{0,\min}]$, with $F_{0,\min} = 60$\,Hz and $F_{0,\max} = 400$\,Hz.

The voicing pseudo-label is obtained by averaging the height of the highest peak over five consecutive frames and thresholding it,
\begin{equation}
u_t = \begin{cases} 1, & \tfrac{1}{5}\textstyle\sum_{i=-2}^{2} \max_{\ell} r_{t+i}[\ell] > 0.5,\\[2pt]
0, & \text{otherwise,} \end{cases}
\label{eq:voice-label}
\end{equation}
which removes isolated errors without delaying the onsets and offsets. The voicing loss is the cross-entropy between the voicing posterior $\hat{u}_t$ of the model and this label,
\begin{equation}
\mathcal{L}_{\text{voice}} = \mathbb{E}_t\Big[-u_t\log \hat{u}_t - (1-u_t)\log(1-\hat{u}_t)\Big].
\label{eq:l-voice}
\end{equation}
The hard voicing decision used for synthesis is detached, so that the other losses cannot reach it through the source mixture, and voicing is trained by $\mathcal{L}_{\text{voice}}$ alone.

The F0 pseudo-label is derived from the autocorrelation of the 0--1.5\,kHz component of the frame, where the harmonics are strongest. A periodic frame has peaks of similar height at every multiple of its period, and the highest of them can fall at twice the period, which would halve the estimate. Instead of the highest peak, the label uses the shortest lag $\ell_t^{\ast}$ whose peak reaches 85\% of the highest one, giving $F_{0,t}^{\ast} = f_s/\ell_t^{\ast}$. The F0 loss is the absolute log-frequency error on the voiced frames $\mathcal{V}$ of the label,
\begin{equation}
\mathcal{L}_{\text{f0}} = \mathbb{E}_{t \in \mathcal{V}}\Big[\big|\log \hat{F}_{0,t} - \log F_{0,t}^{\ast}\big|\Big].
\label{eq:l-f0}
\end{equation}

\begin{figure}[t]
\centering
\includegraphics[width=\columnwidth]{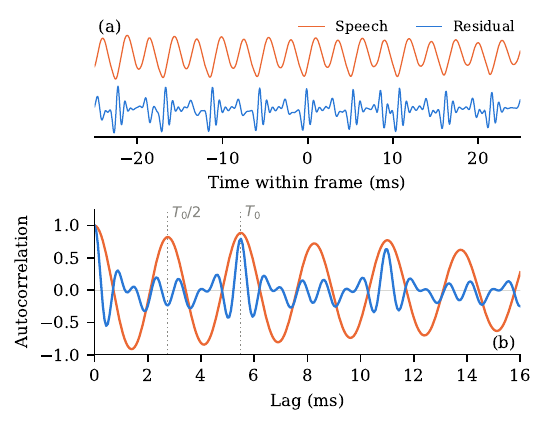}
\caption{A frame of an unseen PTDB-TUG speaker (F04, sa1, 4.38\,s; reference F0 182\,Hz, F1 358\,Hz). (a) The 50\,ms speech frame and its residual after inverse filtering with the estimated filter, both low-pass filtered at 1.5\,kHz. (b) Their normalized autocorrelations; $T_0$ is the reference period.}
\label{fig:resid-acf}
\end{figure}

The autocorrelation of speech is influenced by the formants as well, and a strong formant near a multiple of F0 can pull the peak to the wrong lag. Instead, a second F0 label is derived from the residual, obtained by inverse filtering the speech with the vocal-tract response estimated by the model itself~\cite{markel1972sift, ananthapadmanabha1979epoch}. Figure~\ref{fig:resid-acf} illustrates this on a frame whose first formant lies near the second harmonic. The autocorrelation of the speech is dominated by the formant, peaking almost as high at half the period as at the period (0.82 against 0.89), and its label is an octave too high. The residual is a train of glottal pulses, whose autocorrelation peaks at the period (0.80) and is negative at half the period, and its label is correct. Since the filter is unreliable early on, training starts from the labels of the speech alone. Following the iterative refinement of pseudo-labels in HuBERT \cite{hsu2021hubert}, the F0 loss then moves smoothly from the speech labels to the residual labels over epochs 20--50, $\mathcal{L}_{\text{f0}} = (1-\alpha)\,\mathcal{L}_{\text{f0}}^{\text{speech}} + \alpha\,\mathcal{L}_{\text{f0}}^{\text{resid}}$, with $\alpha$ rising from 0 to 1. The voicing label is always taken from the speech, and no gradient flows through the residual into the filter.

\subsubsection{Waveform reconstruction losses}
Every loss introduced so far compares one frame at a time, so none of them checks whether the per-frame estimates join into a continuous signal. Waveform reconstruction losses fill this gap: by comparing the reconstructed signal with the input sample by sample, they are sensitive to inconsistencies across frames that a per-frame comparison cannot detect.

Let $\hat{s}[n]$ denote the model's reconstructed waveform, obtained by pairing the predicted magnitude spectrum with the phase of the input and inverting. Because the phase is exact, not modeled, $\hat{s}[n]$ stays anchored to the timing of the input even when the predicted parameters are moderately wrong, and a sample-level comparison gives a more reliable gradient than a spectral one. We use an L1 loss and a gain-invariant correlation loss between the reconstructed and input waveforms, given by
\begin{equation}
\mathcal{L}_{\text{wave}} = \frac{1}{N}\sum_n \big|\hat{s}[n]-s[n]\big|
\label{eq:l-wave}
\end{equation}
\begin{equation}
\mathcal{L}_{\text{corr}} = 1 - \frac{\sum_n \hat{s}[n]\,s[n]}{\|\hat{s}\|_2 \, \|s\|_2}
\label{eq:l-corr}
\end{equation}

Together, these four terms define the time-domain loss
\begin{equation}
\mathcal{L}_{\text{time}} = \mathcal{L}_{\text{voice}} + \mathcal{L}_{\text{f0}} + \mathcal{L}_{\text{wave}} + \mathcal{L}_{\text{corr}}
\label{eq:l-time}
\end{equation}

\subsection{Regularization losses}
\label{sec:reg-losses}

\subsubsection{Gate load-balancing loss for the multi-resolution encoder}

Left unattended, the two gates of the encoder (Section~\ref{sec:encoder}) may collapse onto a single resolution and leave the others unused. We therefore penalize uneven use of the resolutions over the batch. Given the batch-mean weights of the source and system gates, $\bar{w}_{\text{source}}$, $\bar{w}_{\text{system}}$, the gate balancing loss is computed as
\begin{equation}
\mathcal{L}_{\text{gate}} = \big\|\bar{w}_{\text{source}}\big\|_2^2 + \big\|\bar{w}_{\text{system}}\big\|_2^2
\label{eq:l-gate}
\end{equation}
Because it acts on the batch mean, individual frames remain free to commit to one resolution.

\subsubsection{Smoothness loss on formant and F0 tracks}
To keep the formant and F0 tracks smooth, we penalize the squared frame-to-frame change of their log-frequencies,
\begin{equation}
\mathcal{L}_{\text{smooth}} = \mathbb{E}_{t,k}\Big[\big(\log F_{k,t+1} - \log F_{k,t}\big)^2\Big],
\label{eq:l-smooth}
\end{equation}
where $k = 0,\dots,4$ indexes F0 and the four formants. Together, these two terms define the regularization loss
\begin{equation}
\mathcal{L}_{\text{reg}} = 0.05\,\mathcal{L}_{\text{gate}} + 0.15\,\mathcal{L}_{\text{smooth}}
\label{eq:l-reg}
\end{equation}

The model is trained to minimize the sum of the losses from the three groups,
\begin{equation}
\mathcal{L} = \mathcal{L}_{\text{freq}} + \mathcal{L}_{\text{time}} + \mathcal{L}_{\text{reg}}
\label{eq:l-total}
\end{equation}
with the weights within each group set empirically on the training data.
\section{Experimental Evaluation and Result Analysis}
\label{sec:evaluation}

We evaluate the proposed SoFilNet on three fronts: formant tracking accuracy against hand-corrected references for test speakers not seen in training, pitch tracking accuracy against laryngograph-derived references on a different corpus, and the physical interpretability of the learned source and filter parameters. The model is trained on the standard TIMIT training set \cite{garofolo1993timit} (4620 utterances, 462 speakers) with no formant or pitch label of any kind. It uses $M_F = 4$ ordered formants, $M_H = 2$ high poles, $M_E = 2$ extra poles and $N = 2$ zeros (Section~\ref{sec:system-decoder}), and is trained for 1000 epochs on 2\,s segments (batch size 16) with Adam (learning rate $2\times10^{-4}$, weight decay $10^{-5}$), a cosine learning-rate schedule decaying to 1\% of its initial value, and gradient-norm clipping. White noise at a randomly drawn SNR is added to the model input only, while the target spectrum is always computed from the clean waveform. All results are reported from the final checkpoint, without any checkpoint selection on labeled data.

\subsection{Formant tracking accuracy}
\label{sec:formant-accuracy}

Formant tracking is evaluated on VTR-FORMANTS \cite{deng2006vtrformants}, which provides hand-corrected F1--F3 tracks for 516 TIMIT utterances. Its test portion is drawn from the TIMIT test set, so it is disjoint in both speakers and utterances from the training data; results are reported on all 192 of its utterances. F4 in VTR-FORMANTS is not hand-corrected \cite{deng2006vtrformants, dissen2019deepformants} and, following prior work, is excluded from all comparisons. The literature reports results on this benchmark in two ways, and we use both. KARMA, DF-RNN, DF-RCNN and FormantNet use the root-mean-square error (RMSE) over all speech segments, excluding only silence. The wider literature uses the mean absolute error (MAE) over vowel segments only \cite{lilley2021formantnet}.

\begin{table}[!t]
\caption{Formant RMSE (Hz) over all speech frames (silence excluded), following Table 3 of \cite{lilley2021formantnet}}
\label{tab:formants-all}
\centering
\begin{tabular}{lccccc}
\hline
Method & Type & F1 & F2 & F3 & Mean \\
\hline
Praat$^\dagger$ & classical & 234 & 232 & 241 & 236 \\
WaveSurfer$^\dagger$ & classical & 171 & 260 & 357 & 263 \\
KARMA (2012) & unsup. & \textbf{114} & 226 & 320 & 220 \\
DF-RNN (2019) & sup. & 118 & 169 & 204 & 163 \\
DF-RCNN (2019) & sup. & 127 & 180 & 213 & 173 \\
FormantNet (2021) & unsup. & 143 & 177 & 195 & 173 \\
\textbf{SoFilNet} & \textbf{unsup.} & 136 & \textbf{156} & \textbf{183} & \textbf{158} \\
\hline
\multicolumn{6}{l}{\footnotesize $^\dagger$Run by us on the same utterances.} \\
\end{tabular}
\end{table}

\begin{table}[!t]
\caption{Formant MAE (Hz) over vowel frames, following Table 2 of \cite{lilley2021formantnet}}
\label{tab:formants-vowel}
\centering
\begin{tabular}{lcccc}
\hline
Method & F1 & F2 & F3 & Mean \\
\hline
Praat$^\dagger$ & 60 & 67 & 105 & 77.2 \\
WaveSurfer$^\dagger$ & 54 & 76 & 115 & 81.8 \\
MSR (2004) & 64 & 105 & 125 & 98 \\
DeepFormants (2019, MLP est.) & 54 & 81 & 112 & 82.3 \\
DF-RCNN (2019) & \textbf{53} & 72 & 108 & 77.7 \\
FormantNet (2021) & 64 & 75 & 90 & 76.3 \\
\textbf{SoFilNet} & \textbf{53} & \textbf{59} & \textbf{88} & \textbf{67.0} \\
\hline
\multicolumn{5}{l}{\footnotesize $^\dagger$Run by us on the same utterances.} \\
\end{tabular}
\end{table}

Tables~\ref{tab:formants-all} and~\ref{tab:formants-vowel} summarize the results. Under both conventions SoFilNet has the lowest mean error of all listed methods, including the supervised DF-RNN and DF-RCNN trackers, although it never sees a formant label. Its advantage lies in F2 and F3, where it is the most accurate method under both conventions. These are the formants most affected by antiresonances and most easily confused with neighboring poles. The zeros of the filter, without which the F2 and F3 errors rise sharply (Section~\ref{sec:ablations}), and the ordering of the formants by construction are aimed at these two difficulties. On F1, KARMA and the supervised trackers perform better over all speech, while on vowels SoFilNet only matches DF-RCNN. F1 lies among the lowest harmonics, where the source and the filter are hardest to separate from the magnitude spectrum alone; FormantNet, also unsupervised, shows the same weakness, and KARMA's temporal tracking and the labels of the supervised trackers presumably help there. The classical trackers Praat and WaveSurfer are competitive with the earlier learned trackers on vowels. Over all speech frames their error is much larger, mainly because of consonants: in fricatives and stop releases, the spectrum is dominated by resonances of the cavity in front of the constriction~\cite{stevens1998acoustic}, which an all-pole tracker follows, whereas the reference continues the vocal-tract resonances smoothly through the consonant, where they are weakly excited or not visible~\cite{deng2006vtrformants}. 

\begin{figure}[t]
\centering
\includegraphics[width=\columnwidth]{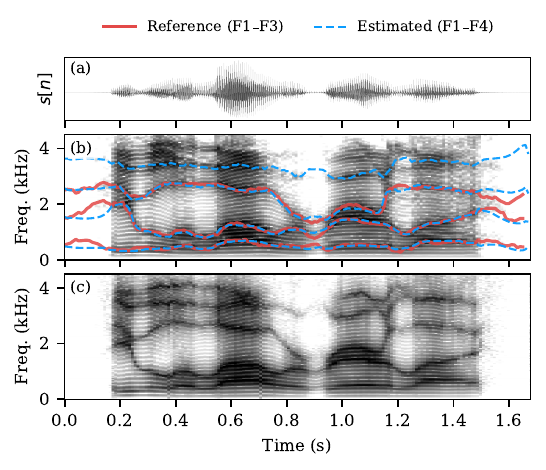}
\caption{A VTR-FORMANTS test utterance (TIMIT test/dr2/mwew0/sx11, ``He will allow a rare lie.''). (a) Waveform. (b) Spectrogram of the speech, with hand-corrected F1--F3 (solid red) and estimated F1--F4 (dashed blue) overlaid; F4 has no hand-corrected reference. (c) Spectrogram predicted from the estimated parameters. Both spectrograms use the 25\,ms analysis window of the model and are pre-emphasized for display.}
\label{fig:diagnostic}
\end{figure}

Figure~\ref{fig:diagnostic} shows the formants estimated by SoFilNet on a test utterance. The estimated F1--F3 follow the reference closely, including through the rapid transitions: the fall of F2 at 0.2--0.3\,s, the joint fall of F2 and F3 at 0.7--0.9\,s and the rise of F3 near 1.15\,s. The spectrogram predicted from the estimated parameters (Figure~\ref{fig:diagnostic}c) reproduces these formant movements together with the harmonic structure of the voiced speech.

\subsection{Pitch tracking accuracy}
\label{sec:pitch-accuracy}

Pitch tracking is evaluated on PTDB-TUG \cite{pirker2011ptdb}, which contains 20 speakers (10 male, 10 female) reading TIMIT prompts, recorded at 48\,kHz in a separate setup. Since the model never sees this corpus, the evaluation is out of domain in both speakers and recording conditions. Its reference F0 is derived from a laryngograph signal (RAPT \cite{talkin1995rapt} on the vocal-fold signal, 10\,ms hop), a physical measurement rather than an estimate from the microphone signal.

We compare against seven pitch trackers, all run on the same 16\,kHz audio: the classical RAPT, Praat \cite{boersma2002praat}, pYIN \cite{mauch2014pyin} and SWIPE \cite{camacho2008swipe}, the supervised CREPE \cite{kim2018crepe}, and the self-supervised PESTO \cite{riou2025pesto} and SPICE \cite{gfeller2020spice}. For each tracker, two settings are chosen by five-fold cross-validation over speakers: the voicing threshold that minimizes the F0 frame error, and a small time lag that aligns its track with the reference. Each of the 20 speakers is thus scored with settings chosen on the other 16. We did not evaluate PENN \cite{morrison2023penn} or SwiftF0 \cite{nieradzik2025swiftf0}, whose released models were trained on PTDB-TUG. We report two groups of metrics. The gross metrics, following the speech literature, count errors at the frame level: the gross pitch error (GPE), the fraction of frames voiced in both reference and estimate whose F0 deviates by more than 20\%; the voicing decision error (VDE), the fraction of all frames with a wrong voicing decision \cite{rabiner1976comparative}; and the F0 frame error (FFE), the fraction of all frames with either error \cite{chu2009ffe}. The fine metrics measure the accuracy of the pitch on every reference-voiced frame, regardless of the voicing decision, as in \cite{riou2025pesto}: raw pitch and chroma accuracy at 50 cents (RPA, RCA), and the mean and median absolute error in semitones (MnE, MdE).

\begin{table}[!t]
\caption{Pitch tracking on all 20 PTDB-TUG speakers. FFE, GPE, VDE, RPA and RCA in \%; MnE and MdE in semitones. $\downarrow$: lower is better; $\uparrow$: higher is better.}
\label{tab:ptdbscore}
\centering
\footnotesize
\setlength{\tabcolsep}{3pt}
\begin{tabular}{lccc|cccc}
\hline
 & \multicolumn{3}{c|}{Gross metrics} & \multicolumn{4}{c}{Fine metrics} \\
Method & FFE$\downarrow$ & GPE$\downarrow$ & VDE$\downarrow$ & RPA$\uparrow$ & RCA$\uparrow$ & MnE$\downarrow$ & MdE$\downarrow$ \\
\hline
\multicolumn{8}{l}{\emph{Classical}} \\
RAPT$^{\ddagger\S}$ & 5.07 & 2.15 & \textbf{4.60} & 81.2 & 82.2 & -- & -- \\
Praat$^\S$ & 5.28 & 2.34 & 4.78 & 84.7 & 86.1 & -- & -- \\
SWIPE$^\S$ & 16.33 & 3.34 & 15.61 & 81.3 & 82.6 & -- & -- \\
pYIN & 8.34 & 4.82 & 7.37 & 76.5 & 78.7 & 1.41 & 0.161 \\
\hline
\multicolumn{8}{l}{\emph{Supervised}} \\
CREPE & 6.70 & \textbf{1.72} & 6.36 & \textbf{85.0} & 87.3 & 0.79 & \textbf{0.141} \\
\hline
\multicolumn{8}{l}{\emph{Self-supervised}} \\
PESTO & 6.13 & 6.13 & 4.81 & 84.5 & \textbf{87.8} & 1.59 & 0.145 \\
SPICE & 13.48 & 2.32 & 13.06 & 73.1 & 75.2 & 1.22 & 0.207 \\
\textbf{SoFilNet} & \textbf{5.06} & 2.12 & 4.61 & 84.4 & 85.4 & \textbf{0.62} & 0.142 \\
\hline
\multicolumn{8}{l}{\footnotesize $^\ddagger$The same algorithm produced the reference.} \\
\multicolumn{8}{l}{\footnotesize $^\S$Pitch estimates not available for all reference-voiced frames.} \\
\end{tabular}
\end{table}

Table~\ref{tab:ptdbscore} summarizes the results. SoFilNet has the lowest F0 frame error, on par with RAPT and well below the supervised CREPE. RAPT is the algorithm that derived the reference from the laryngograph signal, which favors it on this corpus. The advantage of SoFilNet comes from doing well on both components of the measure at once: its voicing decision is as reliable as that of the best classical trackers, and it makes few gross pitch errors. PESTO, also self-supervised, matches SoFilNet in raw pitch accuracy but makes nearly three times as many gross pitch errors. Many of these are octave errors, as the gap between its chroma and raw pitch accuracy shows. Its equivariance objective constrains only relative pitch, so nothing in its training ties each frame to the correct octave. SoFilNet avoids this through its pitch pseudo-labels, derived from the autocorrelation and refined on the residual (Figure~\ref{fig:resid-acf}). Its own misses are mostly small deviations beyond 50 cents, which lower its chroma accuracy below that of PESTO. SPICE falls behind mainly through its voicing decision. CREPE, trained on annotated and synthetic pitch, still makes the fewest gross errors and is slightly more accurate on voiced frames, so supervision retains an edge in fine pitch accuracy. RAPT, Praat and SWIPE give no pitch on the frames they judge unvoiced, which lowers their raw pitch accuracy.

\begin{figure}[t]
\centering
\includegraphics[width=\columnwidth]{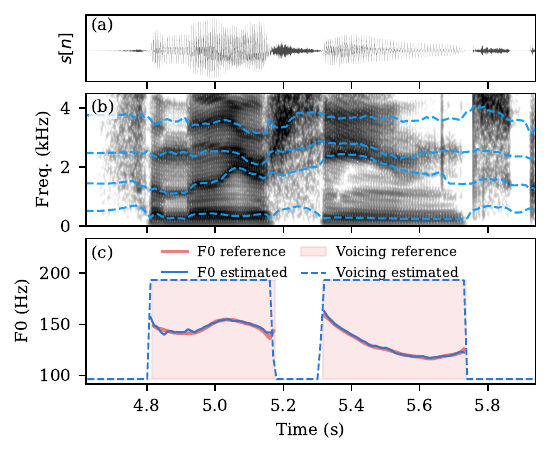}
\caption{A PTDB-TUG utterance (MALE/M07/si1722), unseen in training. (a) Waveform. (b) Wideband spectrogram (12\,ms window), pre-emphasized for display, with estimated formants overlaid (dashed blue; no formant reference exists for this corpus). (c) Estimated F0 (blue), shown on the frames the model classifies as voiced, and estimated voicing (dashed blue) against the laryngograph-derived reference (red), for two voiced segments (4.82--5.17\,s and 5.32--5.73\,s) separated by an unvoiced gap.}
\label{fig:ptdb-example}
\end{figure}

Figure~\ref{fig:ptdb-example} shows a PTDB-TUG utterance. The estimated F0 follows the reference contour within each voiced segment, and the voicing decision switches cleanly at every onset and offset. The formants are also tracked closely, even though the utterance comes from a different corpus, which illustrates the generalization of the proposed approach. Each estimated formant lies on a visible resonance band of the spectrogram and follows it through the transitions: in the first voiced segment F2 rises from about 1.2 to 1.8\,kHz while F3 falls towards it, and in the second F1 stays near 250\,Hz while F2 falls from 2.4 to 1.7\,kHz, with no jumps between formants. Since no formant reference exists for PTDB-TUG, this evidence is qualitative.

\subsection{Interpretability of learned parameters}
\label{sec:interpretability}

Beyond formants and F0, the model estimates the aperiodicity of five bands and two antiresonance zeros. Neither receives a phonetic label: aperiodicity is trained through a proxy, the periodic fraction of each band of the input, and the zeros only through the reconstruction of the spectrum. We examine whether their estimates match the behavior expected from speech production theory. To this end, the frames of the 192 evaluation utterances are grouped by their TIMIT phone label. These frames come from continuous, coarticulated speech rather than from sustained or carefully produced phones, so the averages below describe pooled statistical relations between the estimated parameters and the phone classes, not the acoustic targets of individual phones. Since aperiodicity acts only on the voiced source (Section~\ref{sec:harmonic-noise}), it is averaged over the frames that are voiced according to both the phone label and the model.

\begin{figure}[t]
\centering
\includegraphics[width=\columnwidth]{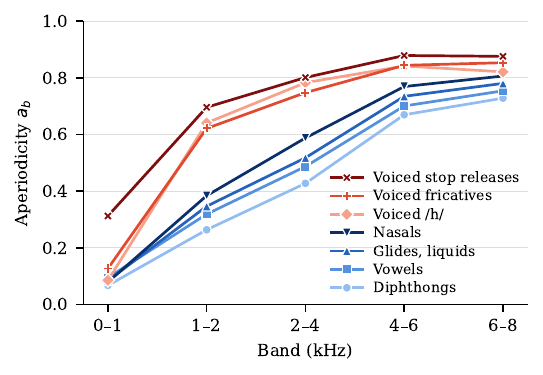}
\caption{Band aperiodicity $a_b$ by broad phone class, averaged over the frames of each class classified as voiced by the model (between 169 frames for voiced /h/ and 15502 for vowels). The legend lists the classes in order of the expected turbulence of their excitation.}
\label{fig:apclass}
\end{figure}

\emph{Aperiodicity rises with frequency and follows the manner of articulation.} In every class, aperiodicity increases from the lowest to the highest band (Figure~\ref{fig:apclass}), as expected of voiced speech, which is mostly harmonic at low frequencies and increasingly noisy at high ones \cite{kawahara2001aperiodicity}. Above 1\,kHz, the classes are ordered by the turbulence of their excitation: lowest for diphthongs and vowels, then glides and liquids and nasals, and highest for voiced /h/, voiced fricatives and voiced stop releases, where voicing is mixed with the noise of aspiration, frication or the stop burst \cite{stevens1998acoustic}. Below 1\,kHz the voiced source dominates, and aperiodicity stays below 0.2 for all classes except the stop releases, where it exceeds 0.3 because the burst is broadband. At 1--2\,kHz, the breathy /h/ is more aperiodic than every class except the stop releases, in agreement with the aspiration noise of a more open glottal configuration \cite{klatt1990analysis}.

\begin{table}[!t]
\caption{Zero radius (depth) and frequency by place of articulation of the nasal consonants. Zero pairs are numbered by frequency.}
\label{tab:zeros}
\centering
\begin{tabular}{lcccccc}
\hline
& \multicolumn{2}{c}{Radius} & \multicolumn{3}{c}{Frequency (Hz)} \\
Pair & Vowel & Nasal & /m/ & /n/ & /\textipa{N}/ \\
\hline
1 & 0.756 & 0.863 & 1149 & 1350 & 1469 \\
2 & 0.597 & 0.780 & 3353 & 3543 & 3514 \\
\hline
\end{tabular}
\end{table}

\begin{figure}[t]
\centering
\includegraphics[width=\columnwidth]{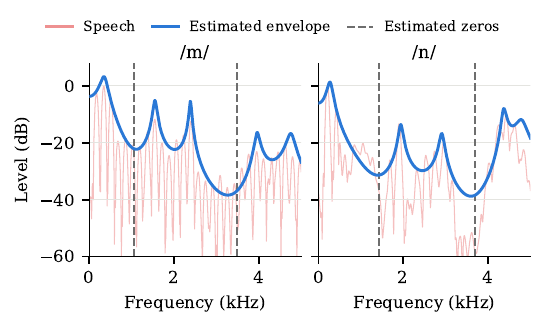}
\caption{Speech spectrum (red) and estimated spectral envelope (blue) in an /m/ (si934) and an /n/ (si1564) of the same test speaker, fmgd0, with the estimated zeros marked (dashed).}
\label{fig:nasal-nulls}
\end{figure}

\emph{Zeros track nasal antiresonances.} Both zero pairs are deeper (larger radius) on nasals than on vowels (Table~\ref{tab:zeros}), as acoustic theory predicts \cite{fant1960acoustic}. The first pair rises with place of articulation from /m/ (1149\,Hz) to /n/ (1350\,Hz) to /\textipa{N}/ (1469\,Hz), as the antiresonance of the closed oral side branch should, since the branch shortens as the closure moves back \cite{fujimura1962nasal}. Its /m/ value lies within the range measured by Fujimura \cite{fujimura1962nasal}, its /n/ value just below that range, and its /\textipa{N}/ value well below the value above 3\,kHz reported for the velar nasal. The second pair lies near 3.4--3.5\,kHz for all three places and is largely confined to nasals: its radius exceeds 0.7 in 79\% of nasal frames but in only 30\% of vowel frames. Figure~\ref{fig:nasal-nulls} compares the two nasals of one speaker: the lower zero moves from 1.05\,kHz for /m/ to 1.42\,kHz for /n/, and the upper one from 3.48 to 3.70\,kHz, each placing a valley in the estimated envelope where the speech spectrum is also weak. Both correspondences emerge without any label for nasality or antiresonance.

\subsection{Reconstruction from interpretable parameters}
\label{sec:reconstruction}

The parameters of Section~\ref{sec:source-filter} describe each 10\,ms frame with 28 numbers: F0, voicing $u$ and five band aperiodicities for the source, the radii and frequencies of eight pole pairs and two zero pairs for the filter, and the gain $\gamma$. To assess how much of the perceptually relevant content of speech these few numbers carry, we resynthesize each evaluation utterance from its estimated parameters alone and measure its quality and intelligibility against the original with wideband PESQ (perceptual evaluation of speech quality) \cite{itu2007pesqwb} and STOI (short-time objective intelligibility) \cite{taal2011stoi}.\footnote{Audio samples available at \url{https://nani-meka.github.io/sofilnet-samples/}}

The synthesis follows the model directly. The harmonic source is a sum of equal-amplitude harmonics below the Nyquist frequency, with a phase that accumulates sample by sample from the predicted F0 track. The aperiodic source is white Gaussian noise. Both are mixed frame by frame in the short-time Fourier domain with the power weights $1-a(\omega)$ and $a(\omega)$ of Equation~\eqref{eq:source-mix}, shaped by the complex response of the predicted pole-zero filter (Equation~\eqref{eq:vz}), which is minimum phase since all poles and zeros lie inside the unit circle, scaled to the frame energy $\gamma$ (Equation~\eqref{eq:combine}), and overlap-added. No phase is taken from the original signal at any point.

\begin{table}[!t]
\caption{Wideband PESQ and STOI of utterances resynthesized from the estimated parameters, of two reference reconstructions, and of the original signal itself (oracle), over the 192 evaluation utterances}
\label{tab:recon}
\centering
\begin{tabular}{llcc}
\hline
Magnitude & Phase & PESQ & STOI \\
\hline
Estimated & Estimated (parameters only) & 2.12 & 0.903 \\
Estimated & Original signal & 3.02 & 0.951 \\
Original signal & Estimated & 3.20 & 0.966 \\
Original signal & Original signal & 4.64 & 1.000 \\
\hline
\end{tabular}
\end{table}

Speech synthesized from the parameters alone reaches a PESQ of $2.12 \pm 0.27$ and a STOI of $0.903 \pm 0.020$ (Table~\ref{tab:recon}, first row). This STOI lies in the range where the measure predicts near-complete intelligibility \cite{taal2011stoi}, so the estimated parameters retain the spectro-temporal information that carries the linguistic content of speech, while the modest PESQ shows that they do not reproduce its full naturalness. The two reference reconstructions show how much each part of the parameters carries. The estimated magnitude, illustrated in Figure~\ref{fig:diagnostic}c, reaches a PESQ of 3.02 when combined with the phase of the original signal, so the filter and source parameters capture most of the spectral information. The phase generated from the estimated F0 track and filter, combined with the original magnitude, reaches 3.20, so the F0 track captures the essential structure of the harmonic phase. These results characterize the perceptual content of the parameters rather than the quality of a speech synthesizer.

\subsection{Ablation studies}
\label{sec:ablations}

The performance of SoFilNet rests on design choices at three levels: the source-filter model, the network architecture and the loss functions. We assess each choice by removing or replacing it, one at a time, and training the resulting variant from scratch under the same conditions (Table~\ref{tab:ablations}).

\begin{table}[!t]
\caption{Ablation results. RMSE (all speech) and MAE (vowels) are F1--F3 means as in Tables~\ref{tab:formants-all} and~\ref{tab:formants-vowel}; FFE and RPA on all 20 PTDB-TUG speakers as in Table~\ref{tab:ptdbscore}.}
\label{tab:ablations}
\centering
\footnotesize
\begin{tabular}{llcccc}
\hline
Group & Ablation & \makecell{RMSE\\(Hz)} & \makecell{MAE\\(Hz)} & \makecell{FFE\\(\%)} & \makecell{RPA\\(\%)} \\
\hline
-- & \textbf{SoFilNet (final)} & 158.3 & 67.0 & 5.06 & 84.4 \\
\hline
Model & No zeros (all-pole) & 245.2 & 92.9 & 5.17 & 82.7 \\
\hline
Arch. & Single shared decoder & 166.3 & 69.2 & 4.92 & 83.0 \\
Arch. & Single-resolution enc. & 165.5 & 68.7 & 5.23 & 85.1 \\
\hline
Loss & No pitch pseudo-labels & 158.0 & 67.6 & 7.60 & 66.8 \\
Loss & No periodic fraction & 163.3 & 66.8 & 5.19 & 83.8 \\
Loss & No contrast losses & 161.0 & 68.8 & 5.11 & 84.0 \\
\hline
\end{tabular}
\end{table}

\emph{Source-filter model.} Removing the zeros reduces the filter to an all-pole model. This variant is by far the worst formant tracker in the table, with an all-speech RMSE of 245.2\,Hz against 158.3\,Hz. F1 is only mildly affected (136.2 to 143.5\,Hz), but F2 rises from 155.6 to 229.2\,Hz and the F3 error doubles (183.1 to 363.0\,Hz). These errors fall back towards those of the classical all-pole trackers in Tables~\ref{tab:formants-all} and~\ref{tab:formants-vowel}, so the zeros account for most of the advantage of SoFilNet on F2 and F3. Without zeros, the poles alone must also reproduce the deep valleys that antiresonances produce, most of all in nasals and laterals. Presumably, they do so by shifting and broadening away from the formants they are meant to represent.

\emph{Network architecture.} In this ablation, we study the contribution of the separate source and system decoders and of the multi-resolution encoder. The first variant replaces the two decoders by a single shared decoder. The second replaces the multi-resolution encoder by a single kernel of size 11, the mean of the kernels 5, 9, 13 and 17, so that the average receptive field is unchanged. The shared decoder raises the all-speech RMSE from 158.3 to 166.3\,Hz, and the single-resolution encoder to 165.5\,Hz, in both cases mainly through F2 and F3. Their effects on pitch are within run-to-run variation: the shared decoder has a slightly lower FFE (4.92\%) but a lower RPA (83.0\%), and the single-resolution encoder the opposite.

\emph{Loss functions.} In this ablation, we study the contribution of the three targets introduced in Section~\ref{sec:losses}: the pitch pseudo-labels, the periodic-fraction loss and the contrast losses. Each variant removes one of them. Without the pitch pseudo-labels, F0 is trained by the reconstruction losses alone. F0 still converges, but makes more than five times as many gross errors, and its raw pitch accuracy falls from 84.4 to 66.8\%. This degradation mainly affects low-pitched voices: the estimate rarely falls below about 100\,Hz, so pitch lower than that is almost never tracked correctly, while above 120\,Hz the accuracy is close to that of SoFilNet. With a 25\,ms analysis window, harmonics spaced less than about 100\,Hz apart are barely resolved, so the spectral losses carry little information about such low F0. On the other hand, the pseudo-labels, computed over 50\,ms, supply this information.

Without the periodic-fraction loss, which is a proxy to train the band aperiodicities, they are trained by the spectral losses like every other parameter. The formant and pitch accuracy degrade slightly, but the source becomes much too periodic. At high frequencies, where the speech is increasingly noisy, the predicted spectrum remains largely periodic, and the estimated aperiodicities fall well below their values with the loss.

Without the spectral-contrast and envelope-extrema losses, the median F1--F3 bandwidths of vowels widen from 109/110/146\,Hz to 126/137/201\,Hz, and the vowel formant error rises from 67.0 to 68.8\,Hz. This is the blurring that per-bin losses produce where formant positions are uncertain. By sharpening the formants, the contrast losses thus contribute a minor improvement.

\subsection{Discussion}
\label{sec:discussion}

SoFilNet combines neural analysis with differentiable synthesis. The synthesis is the classical linear model of speech production, in which a source excites a time-varying pole-zero filter, while its parameters are estimated by a nonlinear network trained to reproduce the speech through this model. The approach thus brings together the estimation power of nonlinear models and the interpretability of linear ones: each parameter is defined by its role in generating the signal rather than by a label. Unlabeled speech is therefore sufficient for training. Estimating the source and the filter jointly is a strength of this design: both must explain the same spectrum, so the harmonic structure of the source is separated from the envelope of the filter, and the formants are estimated as continuous parameters of the model rather than picked as spectral peaks frame by frame. The filter also includes zeros, which model the antiresonances of nasals and laterals. Without them, the poles would have to imitate these spectral valleys and drift away from the formants; with them, the poles stay on the resonances, and the antiresonances themselves become measurable.

A good reconstruction alone, however, does not make every parameter physically meaningful. Different combinations of parameters can explain the same spectrum, and a per-bin spectral loss gives little guidance on F0, drives the aperiodicity too low and blurs the formants. SoFilNet addresses each of these with a target derived from the signal itself: pseudo-labels from the autocorrelation for F0, the periodic fraction of each band as a proxy for aperiodicity, and the spectral contrast for the sharpness of the formants. These targets require no annotation, so the model remains self-supervised, and they make the estimated parameters behave as acoustic phonetics predicts without compromising accuracy.

Because its parameters are those of speech production, SoFilNet can serve several fields. In phonetics, it estimates formants, F0, voicing and aperiodicity in a single pass, without the speaker-specific settings that classical analysis tools require \cite{schiel2018formant}, which would ease measurement on large corpora. In clinical speech analysis, its parameters correspond to quantities that clinicians already assess: band aperiodicity relates to breathiness and hoarseness, and the formant tracks to the vowel-space and transition measures used for motor speech disorders. In speech synthesis and voice conversion, the parameters can serve as explicit controls, for example to shift formants, change F0 or increase breathiness before resynthesis, or as conditioning for a neural vocoder. Each of these uses will require validation beyond the healthy adult read speech studied here, in particular on pathological, child and noisy speech.

\subsection{Limitations}
\label{sec:limitations}

Self-supervised methods have so far mostly addressed formants or pitch separately. This work takes a small but significant step towards a unified model of speech production: it recovers the source and the filter together from raw speech. Its compact source-filter structure leaves room to capture further characteristics of speech production, which we outline here as directions for future work.

The model is realized in the frequency domain. It describes each 10\,ms frame by its magnitude spectrum, and the excitation only through spectral properties: F0, voicing and band aperiodicity. It therefore does not capture the time-domain events of phonation. The glottal closure instants, at which the vocal tract receives its main excitation in each cycle \cite{murty2008epoch}, are not estimated. Nor is the shape of the glottal flow pulse, conventionally described by models such as the Liljencrants--Fant model \cite{fant1985lf}. Cycle-to-cycle variations of period and amplitude, known as jitter and shimmer, are averaged within each frame. The source also has no spectral tilt, the frequency-domain counterpart of the pulse shape. In a magnitude-only model, a source tilt competes with the aperiodicities and with the slope of the filter, so the slope is left to the filter, as in linear prediction, and voice-quality differences in spectral slope are not represented separately. Recovering these properties requires a time-domain model of the excitation, such as a parametric glottal pulse driving the estimated filter.

The model also treats source and filter as independent. It cannot represent source-tract interaction, in which the vocal-tract load shapes the glottal flow and the formant bandwidths change between the open and closed phases of the glottal cycle. The coupling of the nasal cavity is represented only through the antiresonances it produces, not through an articulatory description of the tract.

Further simplifications are built into the signal model. Each frame is assumed stationary over its 25\,ms analysis window, so rapid changes within a frame, such as those at stop releases, are averaged. The aperiodic component is stationary noise whose level is set per band, whereas aspiration noise in voiced speech is modulated in synchrony with the glottal cycle, and five fixed bands limit how finely its spectral distribution can vary. The filter has a fixed number of poles and zeros in fixed roles, so frames that need more antiresonances, or resonances outside the frequency range allowed for each pole, can only be approximated. These simplifications lie in the synthesizer rather than in the approach itself. Each can be relaxed by enriching the synthesizer, while the core principle remains the same: physically defined parameters are learned by reproducing the speech through it.

\section{Conclusion}
\label{sec:conclusion}

We presented SoFilNet, a fully differentiable, self-supervised source-filter model that estimates a physically interpretable description of speech production directly from the raw waveform, without formant or pitch labels. On the source side, it estimates F0, voicing and band aperiodicity; on the filter side, ordered formants, additional envelope-shaping poles and antiresonance zeros. A model trained by reconstruction alone tends to blur its estimates, because per-bin spectral losses favor smooth spectra wherever the model is uncertain. SoFilNet complements reconstruction with objectives that capture the periodicity and spectral sharpness that per-bin losses miss. It tracks formants with a lower mean error than supervised trackers, and F0 with a lower frame error than other learned trackers on an unseen corpus, while its aperiodicities and zeros follow speech-production phenomena that were never training targets. More broadly, these results show that the classical model of speech production can be learned from unlabeled speech when the training objective ties each parameter to its physical role, an approach we expect to extend to richer models of speech production, such as those with a time-domain glottal excitation.

\bibliographystyle{IEEEtran}
\bibliography{references}

\end{document}